\documentclass[10pt,journal]{IEEEtran}

\usepackage[pdftex]{graphicx}
\DeclareGraphicsExtensions{.pdf,.jpeg,.png}

\usepackage{multirow,setspace,verbatim,amsfonts,graphicx,amsmath,amsthm,amsbsy,amssymb,epsfig}
\usepackage{url,cite}
\usepackage{booktabs}

\usepackage{array}
\usepackage{booktabs}

\begin{document}

\title{RAFP-Pred:  Robust Prediction of Antifreeze Proteins using Localized Analysis of n-Peptide Compositions}
\author{Shujaat~Khan, Imran~Naseem, 
	Roberto~Togneri,~\IEEEmembership{Senior~Member,~IEEE}
	and~Mohammed~Bennamoun,~\IEEEmembership{Senior~Member,~IEEE} 
	\thanks{Shujaat Khan is with the Faculty of Engineering Science and Technology, Iqra University, Defence View, Shaheed-e-Millat Road (Ext.) 
		Karachi-75500, Pakistan.  Email: shujaat@iqra.edu.pk}
	\thanks{Imran Naseem and Roberto Togneri are with the School of Electrical, Electronic and Computer Engineering, The University of Western Australia, 35 Stirling Highway, Crawley, WA 6009, Australia.  Email: \{imran.naseem,roberto.togneri\}@uwa.edu.au}
	\thanks{Mohammed Bennamoun is with the School of Computer Science and Software Engineering, The University of Western Australia, 35 Stirling Highway, Crawley, WA 6009, Australia.  Email: mohammed.bennamoun@uwa.edu.au. }
	\thanks{This work was published in IEEE/ACM Transactions on Computational Biology and Bioinformatics (TCBB) \cite{RAFP_original}. Code and dataset is available at https://goo.gl/3i7gQD.}}

\maketitle

\begin{abstract}
In extreme cold weather, living organisms produce Antifreeze Proteins (AFPs) to counter the otherwise lethal intracellular formation of ice.  Structures and sequences of various AFPs exhibit a high degree of heterogeneity, consequently the prediction of the AFPs is considered to be a challenging task.  In this research, we propose to handle this arduous manifold learning task using the notion of localized processing.  In particular an AFP sequence is segmented into two sub-segments each of which is analyzed for amino acid and di-peptide compositions.  We propose to use only the most significant features using the concept of information gain (IG) followed by a random forest classification approach.  The proposed RAFP-Pred achieved an excellent performance on a number of standard datasets.  We report a high Youden's index (sensitivity+specificity-1) value of 0.75 on the standard independent test data set outperforming the AFP-PseAAC, AFP\_PSSM, AFP-Pred and iAFP by a margin of 0.05, 0.06, 0.14 and 0.68 respectively.  The verification rate on the UniProKB dataset is found to be 83.19\% which is substantially superior to the 57.18\% reported for the iAFP method.
\end{abstract}


\IEEEpeerreviewmaketitle

\section{Introduction}
%
\IEEEPARstart{I}ce has an unusual property called recrystallization.  When water starts to freeze, it forms many small crystals.  Some of the small crystals soon dominate and continue to become large by stealing water molecules from the surrounding small crystals \cite{yu2001winter}.  This phenomenon can prove to be particularly lethal for living organisms in extreme cold weather due to the intracellular formation of ice \cite{griffith1997antifreeze}.  Antifreeze proteins (AFPs) neutralize this recrystallization effect by binding to the surface of the small ice crystals and retarding the growth into larger dangerous crystals \cite{davies2002structure}\cite{fletcher2001antifreeze}.  Therefore they are also called as 'ice structuring proteins' (ISPs).  The AFPs lower the freezing point of water without altering the melting point, this interesting property of the AFPs is called as 'thermal hysteresis' \cite{urrutia1992plant}.  

The AFPs are critical for the survival of living organisms in extremely cold environments.  They are found in various insects, fish, bacteria, fungi and overwintering plants such as gymnosperms, ferns, monocotyledonous, and angiosperms \cite{yu2001winter},\cite{davies2002structure},\cite{urrutia1992plant},\cite{scholander1957supercooling}, \cite{moriyama1995seasonal},\cite{logsdon1997origin},\cite{ewart1999structure},\cite{cheng1998evolution},\cite{davies1997antifreeze}.  Several studies on various AFPs have shown that there is little structural and sequential similarity for an ice-binding domain \cite{davies2002structure}.  This inconsistency relates to the lack of common features in different AFPs and therefore a reliable prediction of AFPs is considered to be an ardent task.      

The Recent success of machine learning algorithms in the area of protein classification, has encourage several researchers to develop automated approaches for the identification of AFPs.  AFP-Pred \cite{kandaswamy} is considered to be the earliest work in this direction.  The work is essentially based on random forest approach making use of the sequence information such as functional groups, physicochemical properties, short peptides and secondary structural element.  In AFP\_PSSM \cite{xiaowei} evolutionary information is used with support vector machine (SVM) classification.  In iAFP \cite{yu} n-peptide composition is used with limited experimental results.  In particular amino acids, di-peptide and tri-peptide compositions were used.  We argue that tri-peptide composition is computationally expensive (require the calculation of $20^3$ combinations) resulting in redundant information.  Consequently the selection of the most significant features using genetic algorithms (GA) has shown limited results \cite{yu}.  It is also worth noting that n-peptide compositions were derived for the whole sequence.  The latest work in this regard is AFP-PseAAC \cite{mondal} where the pseudo amino acid composition is used with an SVM classifier to achieve a 'good' prediction accuracy.  

In machine learning, the difficult manifold learning problems can effectively be addressed using a localized processing approach compared to its holistic counterparts \cite{kovnatsky2015madmm}.  Considering the diversified structures of AFPs, it is intriguing to explore the localized processing of the protein sequences.  We therefore propose to adopt a segmentation approach where each protein sequence is segmented into two sub-sequences.  The amino acids and di-peptide compositions are derived for each sub-sequence from which we extract the relevant features.  The most significant features are further selected using the concept of information gain and the random forest approach is used for classification.  To the best of our knowledge, this is the first time that localized processing is proposed to deal with the challenging problem of learning diversified structures of the AFPs.  The proposed method has shown to comprehensively outperform all the existing approaches on standard datasets(section 3).          

The paper is organized as follows:  the details and mathematical framework of the proposed approach is presented in Section \ref{proposed}, followed by the description of the data sets and our experimental results in Section \ref{secresult}. Our conclusions are provided in Section \ref{conclusion}.

\section{Proposed Approach}\label{proposed}

The reliable prediction of proteins can only be achieved by robustly encoding the protein sequences into mathematical expressions.  This ensures that the underlying structures of the protein sequences have been truly learned.  In the absence of robust learning methods of the protein sequences, the predictor is unlikely to perform well for unseen test samples.  From the machine learning perspective, the difficult manifold learning problems are effectively tackled using the localized processing approach \cite{kovnatsky2015madmm,yang2008prediction}.  While holistic methods deal with the training samples in a global sense, the localized learning focuses on the various segments of the samples.  Typically, features extracted from confined segments are efficiently fused.  For the challenging manifold learning problems, localized learning has shown to outperform its counterparts in various applications of machine learning \cite{kandaswamy2013ecmpred,dehzangi2015gram,zhang2008using}.  We therefore propose a local analysis approach of AFPs for feature extraction.

\subsection{Features}\label{proposed1}
Since the structures of various AFPs are uncorrelated and lack in similarity, the automated prediction of AFPs is therefore considered to be a challenging task.  Motivated by the robustness of the localized learning approaches, we propose an approach that processes the localized segments of the AFP sequences.  In particular, each protein sequence is segmented into two sub-sequences, each sub-sequence is individually analyzed for amino acid and di-peptide compositions.  

Consider a protein chain of $L$ amino acid residues:

\begin{eqnarray}
	\mathbf{P}&=&R_1R_2R_3\ldots R_L
\end{eqnarray}

where $R_i$ represents the $i^{th}$ residue of protein $\mathbf{P}$ \cite{chou2011some}.  According to the amino acid composition protein $\mathbf{P}$ can be expressed as an array of occurrence frequency of the twenty native amino acids:

\begin{eqnarray}
	\mathbf{P}&=&[f_1f_2f_3\ldots f_{20}]^{T}
\end{eqnarray}

where $f_j;\ j=1,2,3,\ldots,20$ is the normalized occurrence frequency of the $j^{th}$ native amino acid in $\mathbf{P}$, and $T$ is the vector transpose operator. Accordingly, the amino acid composition of a protein can readily be derived once the protein sequencing information is known. This simple, but effective, amino acid composition (AAC) model has been widely used in a number of statistical methods to predict protein structures \cite{horton2007wolf}, \cite{du2012pseaac}.

Dipeptide compositions are computed using 400 ($20\times20$) dipeptides, i.e. AA, AC, AD,$\ldots$, YV, YW, YY.  Each component is calculated using the following equation:

\begin{multline}
	\mbox{fraction of the }\ k^{th}\ \mbox{dipeptide} \\
	= \frac{\mbox{total number of the}\ k^{th}\ \mbox{dipeptide}}{\mbox{total number of all possible dipeptides}}
\end{multline}

\begin{table}
	\caption{List of features}
	\begin{tabular}{ll}
		\hline
		\textbf{Features} & \textbf{Number of attributes} \\
		\hline
		Segment 1    &        \\
		\hline
		Amino Acid Composition features &         20 \\
		\hline
		Dipeptide Composition features &        400 \\
		\hline
		Segment 2 &             \\
		\hline
		Amino Acid Composition features &         20 \\
		\hline
		Dipeptide Composition features &        400 \\
		\hline
		Total &        840 \\
		\hline
	\end{tabular}
	\label{features} 
\end{table}

The 20 AACs and 400 dipeptide compositions are combined to form 420 attributes for each segment of the AFP sequence.  Finally the 420 attributes of individual sub-sequences are fused to form a single representative feature vector consisting of 840 attributes.  Table \ref{features} shows a list of derived features. 

It is well established that the redundant information tends to degrade the classification results \cite{ding2005minimum}.  It is therefore customary to select the most relevant features for the purpose of  classification \cite{koller1996toward}, \cite{langley1994selection}.  Information gain (IG) or Info-Gain is considered to be an important criterion for the selection of the most significant features \cite{kandaswamy2010spred}.  Given a training set $S$ and an attribute $A$, the information gain with respect to the attribute $A$, can be defined as a reduction in entropy of the training set once the attribute $A$ is observed \cite{mitchell1997machine}, mathematically:
\begin{eqnarray}
	IG(S,A)=H(S)-H(S/A)       
\end{eqnarray}

where $H(S)$ is the entropy of $S$ and $H(S/A)$ is the entropy of $S$ conditioned to the observation of attribute $A$.  For the classical case of a dichotomizer:

\begin{eqnarray}
	H(S)=-\sum_{l=1}^{2}p_{l}\log_{2}p_{l}
\end{eqnarray}

and 

\begin{eqnarray}
	H(S/A)=\sum_{v\in Values(A)}\frac{|S_v|}{|S|}H(S_v)
\end{eqnarray}

where $Values(A)$ is a set of all possible values of the attribute $A$,  $S_v$ is the partition of the training set characterizing the value $v$ of attribute $A$, $H(S_v)$ is the entropy of $S_v$ and $|.|$ is the cardinality operator \cite{mitchell1997machine}.

We propose to use the concept of the Info-Gain for the selection of the most significant features from a pool of 840 features (discussed in Section \ref{proposed1}). The features are ranked using the above formulation of IG in a descending order such that the attribute with the highest IG is given the top priority.

\subsection{Classification}
The Random forest approach has shown to produce excellent results for various prediction problems in proteomics \cite{kandaswamy2010spred,kandaswamy2013ecmpred,kandaswamy,wu2003comparison,lee2005extensive,diaz2006gene,kumar2009dna,masso2010knowledge}.  Random forest is an ensemble classification protocol which combines several weak classifiers (decision trees) to constitute a single strong classifier.  The decision trees generated by the random forest approach are combined using a weighted average scheme \cite{breiman2001random}.  The approach harnesses the power of many decision trees, rational randomization, and ensemble learning to develop accurate classification models \cite{breiman2001random}. 

\begin{figure}[h!]
	\begin{center}
		\centering
		\includegraphics[width=8cm]{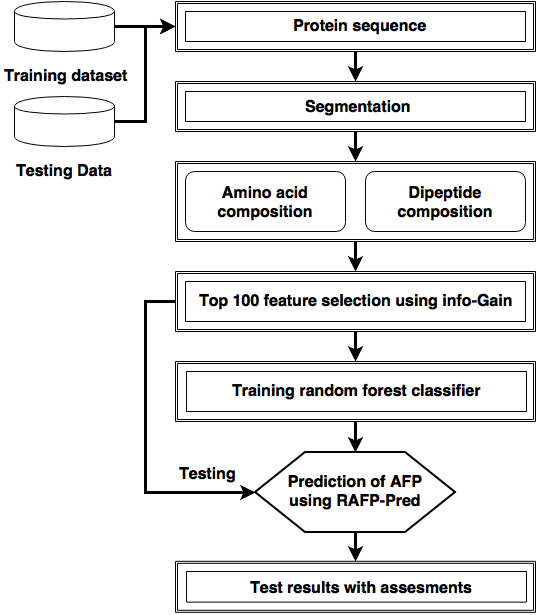}
	\end{center}
	\caption{Work-flow of the proposed RAFP-Pred approach.}
	\label{WorkFlow1}
\end{figure}

Random forest is a supervised learning approach consisting of two steps:  (1) bagging, and (2) random partitioning.  In bagging several decision trees are grown by drawing multiple samples (with replacement) from the original training data set.  Although an indefinite number of such trees can be grown, typically 200-500 trees are considered to be enough \cite{mitchell1997machine}.  The Random forest approach introduces randomness in tree-growing by first randomly selecting a subset of prospective predictors and then producing the split by selecting the best available splitter.  The approach is robust to overfitting and quite efficient on large datasets \cite{breiman2001random}.  A Random forest classifier was implemented using the WEKA tool \cite{frank2004data}, with the following controlling parameters: (1) maximum depth of tree = 10, (2) number of features = 100, (3) number of trees = 50, and (4) number of seeds = 1.  The work-flow of the proposed RAFP-Pred is shown in Figure \ref{WorkFlow1} 	

\section{Experimental Results} \label{secresult}

\subsection{Evaluation Parameters}
For any prediction framework, the Receiver Operating Characteristic (ROC) is considered to be the most comprehensive performance criterion.  The proposed algorithm was therefore extensively evaluated for true positive rate (sensitivity), true negative rate (specificity), prediction accuracy and the area under the curve (AUC).  The proposed algorithm was also evaluated for Matthew's Correlation Coefficient (MCC).  MCC ranges from -1 to 1 with values of MCC = 1 and MCC = -1 indicating the best and the worst predictions respectively,  MCC = 0 shows the case of a random guess.  Youden's index (or Youden's J statistics) is an interesting way of summarizing the results of a diagnostic experiment \cite{youden1950index}.  Ranging from 0 to 1, 0 indicates the worst performance while 1 shows perfect results with no false positives and no false negatives.  Youden's index is typically useful for the evaluation of highly imbalanced test data. 

\subsection{Experimental Results}
Extensive experiments were conducted on a number of state-of-the-art datasets reported frequently in the literature \cite{kandaswamy}, \cite{yu}.

\subsubsection{Dataset 1} \label{dataset1} Dataset 1 consists of 481 AFPs and 9493 non-AFPs reported in \cite{kandaswamy}.  The dataset is further partitioned into training and testing sets.  The training set characterizes 300 AFPs and 300 non-AFPs selected randomly from a pool of 481 AFPs and 9493 non-AFPs respectively.  The remaining 181 AFPs and 9193 non-AFPs constitute the testing set.  Training accuracy was achieved by evaluating the proposed algorithm on the 600 training samples.  This dataset is obtained from \cite{kandaswamy} in which the protein sequences were collected from the Pfam database \cite{ sonnhammer1997pfam}.  For the redundancy check the PSI-BLAST search was performed for each sequence against a non-redundant sequence database with a stringent threshold (E-value 0.001) and followed by the manual inspection to retain only antifreeze proteins. The final dataset contains only the protein sequence with $<=$40\% sequence and all other similar proteins were removed from the dataset using CD-HIT \cite{li2001clustering}.

The proposed approach attained 100\% accuracy on a randomly selected training set which outperforms the AFP-Pred method by a margin of 18.67\% \cite{kandaswamy} and the AFP\_PSSM method by a margin of 17.33\% \cite{xiaowei}.  The average accuracy of three randomly selected training sets, for the proposed method, was found to be 99.91\% with a standard deviation of 0.16\%.  This prediction performance is 10.22\% better compared to the AFP-PseAAC approach (standard deviation of 0.706\%)\cite{mondal}.  
\begin{table*}
	\begin{center}
		\caption{Performance of the proposed RAFP-Pred on test dataset containing 181 AFPs and 9193
			non-AFPs using different feature subsets.}
		\begin{tabular}{cccccc}
			\hline
			{\bf Feature subset} & {\bf Sensitivity (\%)} & {\bf Specificity (\%)} &  {\bf MCC} & {\bf Accuracy (\%)} & {\bf Youden's index} \\
			\hline
			\\
			25 features &    79.01\% &    89.24\% & 0.288 &    89.04\% & 0.68 \\
			\hline
			50 features &    82.32\% &    90.03\% & 0.314 &    89.88\% & 0.72 \\
			\hline
			75 features &    81.77\% &    89.83\% & 0.308 &    89.67\% & 0.72 \\
			\hline
			{\bf 100 features} & {\bf 83.98\%} & {\bf 91.07\%} & {\bf 0.339} & {\bf 90.93\%} & {\bf 0.75} \\
			\hline
			200 features &    79.01\% &    90.10\% & 0.301 &    89.88\% & 0.69\\
			\hline
			400 features &    80.11\% &    90.93\% & 0.320 &    90.72\% & 0.71 \\
			\hline
			600 features &    82.87\% &    90.20\% & 0.319 &    90.06\% &  0.73 \\
			\hline
			800 features &    82.87\% &    89.67\% & 0.310 &    89.54\% & 0.72 \\
			\hline
			All features &    83.43\% &    89.22\% & 0.306 &    89.11\% & 0.73 \\
			\\
			\hline
		\end{tabular} 
		\label{testdata1}
	\end{center}
\end{table*}

The results for the test data set, using different feature subsets, are shown in Table \ref{testdata1}.  The proposed RAFP-Pred achieves the best accuracy of 90.93\% utilizing the 100 most significant features.  For a comprehensive evaluation, the proposed approach was also compared to the state-of-art methods reported in the literature (refer to Table \ref{testdata1_1}).  Note that for a fair comparison we implemented and evaluated all approaches using the same training and testing examples.  We were however unable to generate results for AFP\_PSSM \cite{xiaowei} as the data was unavailable during our experiments.  Instead we compared our results directly with those reported in \cite{xiaowei}.    


\begin{table*}
	\begin{center}
		\caption{Comparison of the proposed RAFP-Pred with different machine learning approaches on Dataset 1.}
		\begin{tabular}{cccccc}
			\hline
			{\bf Predictor} & {\bf Sensitivity (\%)} & {\bf Specificity (\%)} &  {\bf Accuracy (\%)} & {\bf Youden's index} & {\bf AUC}   \\
			\hline
			\\
			iAFP  &     9.94\% &    97.23\%  &    95.55\% &     0.07 & NA \\ 
			\hline
			AFP-Pred &    82.32\% &    79.02\%  &    79.08\% &     0.61 & 0.89\\
			\hline
			AFP\_PSSM \cite{xiaowei} &    75.89\% &    93.28\%  &    93.01\% &     0.69 & 0.93\\
			\hline
			AFP-PseAAC  &    82.87\% &    87.61\%  &    87.52\% &     0.70 & NA\\
			\hline
			{\bf RAFP-Pred} & {\bf 83.98\%} & {\bf 91.07\%} & {\bf 90.93\%} & {\bf 0.75} & 0.95 \\
			\\
			\hline
		\end{tabular} 
		\label{testdata1_1}
	\end{center}
\end{table*}

The test data set is highly imbalanced with 181 (AFPs) positive and 9193 (non-AFPs) negative examples.  For such a highly imbalanced test data, there is a natural tendency for the predictor to be biased in favor of the class which has more samples.  In such scenarios, the evaluation parameters such as the AUC and Youden's index are more representative of the predictor's performance than the conventional sensitivity, specificity and accuracy measures. 

\begin{figure}[h!]
	\begin{center}
		\centering
		\includegraphics[width=8cm]{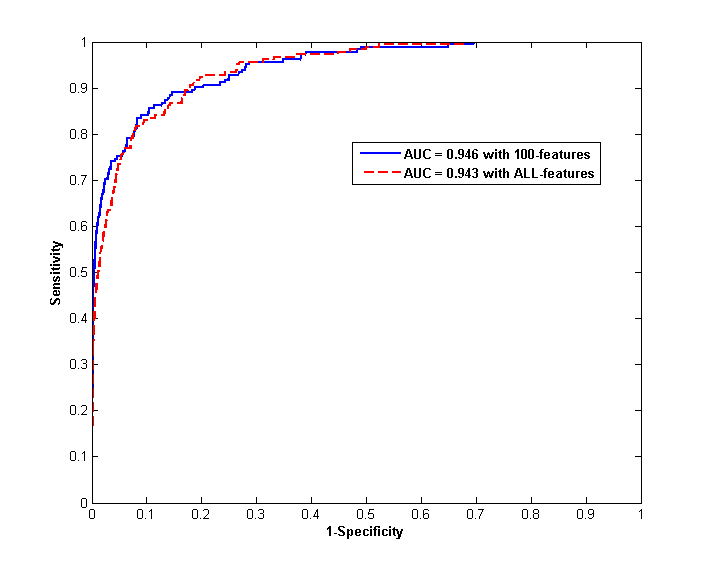}
	\end{center}
	\caption{ROC curves for the proposed RAFP-Pred approach.}
	\label{roc}
\end{figure}

 For instance in Table \ref{testdata1_1} iAFP achieves a very high specificity of 97.23\% but a poor sensitivity of 9.94\%. Therefore, although the overall accuracy of 95.55\% appears to be the best reported accuracy, the predictor has a low Youden's index of 0.07 and therefore cannot be regarded as competitive. The proposed approach achieved a Youden's index of 0.75 which is better than all reported results in the literature.  The receiver operating characteristics (ROC) are shown in Figure \ref{roc} where the highest AUC of 0.95 verifies the excellent performance of the proposed RAFP-Pred approach. 
 
  The 100 most significant features obtained using the training samples of dataset 1 are available online at https://goo.gl/3i7gQD.  These 100 features were used for all the datasets. 

It is interesting to compare the proposed approach with the latest and the most successful method reported in literature i.e., AFP-PseAAC.  The proposed RAFP-Pred approach has shown to comprehensively outperform the AFP-PseAAC method.  The AFP-PseAAC achieved a sensitivity and specificity of 82.87\% and 87.61\% respectively which lags the proposed approach by a margin of 1.11\% and 3.46\%.  The Youden's index of the AFP-PseAAC was also found to be 0.70 which is inferior to the proposed approach.

\subsubsection{Dataset 2}  Dataset 2 consists of 44 AFPs and 3762 non-AFPs collected from the Protein Data Bank (PDB) \cite{berman2000protein} and the PISCES server \cite{wang2003pisces} respectively (reported in \cite{yu}).  The non-AFPs in the dataset had, 25\% pairwise sequence identity (SI), R-factors of 0.25 and a crystallographic resolution of at least 2 A$^{0}$.  In this dataset only those AFPs that had known 3D structures are included. Dataset 2 is also a highly imbalanced dataset with 44 positive and 3762 negative examples.  In the literature, the only results reported on this dataset are for the iAFP method \cite{yu}.  In particular the iAFP method attained an accuracy of 99.32\% on 7-fold cross validation.  The proposed RAFP-Pred approach attained a comparable accuracy of 99.71\% using the 100 most significant features obtained in section \ref{dataset1}.  The MCC value of 0.87 found for the proposed RAFP-Pred is also favorably comparable to the 0.79 reported for the iAFP method.  Note that the proposed RAFP-Pred approach was trained using the samples of dataset 2 only and no other training samples were used.  For each iteration of 7-fold cross validation, the redundancy between the training and testing samples was explicitly checked and all samples were found to be unique.  


The state-of-art AFP-PseAAC approach achieved an accuracy of 99.74\% which is quite comparable to the 99.71\% of the proposed approach.  The MCC value of 0.88 achieved by the AFP-PseAAC is also comparable to the 0.87 attained by the proposed RAFP-Pred approach.

\subsubsection{Dataset 3}  Dataset 3 is an independent dataset representing an evolutionarily divergent group of organisms consisting of 369 AFPs obtained from the UniProKB database by searching for the phrase ``antifreeze" \cite{bairoch2000swiss}, \cite{uniprot2010universal}.  Any redundancies i.e., duplicate sequence or partial sequences, were removed during the search. To further filter the dataset all sequences were also removed that were labeled as ``predicted" and ``putative" in the protein name field and followed by the manual check against the literature.  To avoid any confusion any proteins which belong to ``antifreeze-like proteins" were also excluded.  The results on this dataset are reported only for the iAFP method \cite{yu} in \cite{mondal}.  The proposed RAFP-Pred was trained using the training data in \cite{mondal} (i.e. dataset 2), where the 100 most significant features were used.  The sequences of training and testing sets were scanned for similar sequences and no identical sequences were found.  The proposed RAFP-Pred approach attained the highest verification of 83.19\% which is substantially better than the 57.18\% reported for the iAFP. 

The AFP-PseAAC approach was also evaluated using the same training and testing samples achieving a verification rate of 40.17\% which is 43.02\% inferior to the proposed RAFP-Pred approach.



\section{Biological Justification of the Most Significant Features Selected by the Proposed Approach}

It is well known that the biological proteins usually have hydrophobic amino acids in the core (away from water molecules in the solvent).  Interestingly, some AFPs have many hydrophobic amino acids on their surfaces \cite{AFPstructure1}, \cite{AFPstructure2}, \cite{AFPstructure3}.  On the other hand $\alpha$-helices are most commonly found at the surface of the protein cores (for the case of some fish AFPs for instance) where they provide an interface with the aqueous environment.

Regions which tend to form an $\alpha$-helix are: (1) Richer in alanine (A), glutamic acid (E), leucine (L), and methionine (M), and (2) poorer in proline (P), glycine (G), tyrosine (Y), and serine (S).  Careful analysis of the localized segments show that:
\begin{itemize} 
	\item Segment 1 contains high Proline, high Serine, high Tyrosine and low Alanine which indicates less likelihood of an $\alpha$-helix in segment 1.
	\item Segment 2 contains low Tyrosine, high Glutamic Acid, high Alanine and moderate Methionine which indicates high probability of an $\alpha$-helix in segment 2.
\end{itemize}
The above discussion shows that segment 2 has a high probability of an $\alpha$-helix region.   Biologically, we can expect AFPs to have more hydrophobic amino acids in segment 2 compared to the non-AFPs.  This can serve as a biologically justified point of discrimination as such. 
The features selected by the proposed RAFP-Pred contains about 68\% of the features from segment 2 and the 58\% of the segment 2 features are hydrophobic amino acid related features.  It therefore follows that the proposed approach selected the most relevant and biologically justified features for the AFP prediction.

The structural and sequential diversity in AFPs demands a feature-set encompassing a broader range of features catering for most types of AFPs.  For instance, the cysteine composition my vary for different organisms, conserved cysteines form disulfide bonds in beta-helix insect AFP but the same is not true for type 1 fish AFPs.  A broader range of features is therefore required to predict AFPs across organisms.  A thorough investigation shows that the optimal feature-set obtained by the proposed RAFP-Pred approach indeed contains a broad spectrum of these significant features. 

For instance type 1 AFPs are rich in alanine amino acid \cite{Type1_AFP}, type 2 and type 5 AFPs are rich in cysteine amino acid \cite{Type2_AFP}, \cite{Type5_AFP} and Type 4 AFPs are rich in glutamine amino acid \cite{Type4_AFP1}, \cite{Type4_AFP2}.  Interestingly the optimal feature set obtained by the proposed RAFP-Pred approach contains all these features.  This explains the better performance of the proposed approach compared to the contemporary predictors.

In our experiments, the training data of dataset 1 (300 AFPs and 300 Non-AFPs) was used to identify the top 100 significant features. The details are provided in the supplementary material.  Here we discuss the top three features selected by the proposed approach.

The most relevant feature selected by the proposed approach is the frequency of the tryptophan amino acid in segment 2.  A careful exploration of the training data shows that segment 2 of the AFPs contains 40.97\% more tryptophan compared to the non-AFPs.  It is therefore safe to assume that the frequency of the tryptophan amino acid in segment 2 is a discriminating feature. The second most relevant feature selected by the proposed approach is the frequency of the leucine amino acid in segment 2.  The non-AFPs of the training data set contains 20.82\% more leucine compared to the AFPs counterpart; leucine can therefore be regarded as another discriminating feature. The frequency of occurrence of the amino acid cysteine is the third most relevant feature that is selected by the proposed approach.  Analysis on the training data shows that it is found in abundance in both segments of the AFPs compared to the non-AFPs.  In particular the training data contained 36.30\% more cysteine in the AFPs than the non-AFPs and therefore cysteine is regarded as an important discriminating feature.  This finding is supported by other researches who highlight the significance of cysteine in the prediction of the AFPs \cite{Type2_AFP}, \cite{Type5_AFP}.  In fact 19 out of 100 features selected by the proposed approach are cysteine related.

For further details on all the selected features, the reader is referred to the supplementary material.

\section{Conclusion}\label{conclusion}

The structural and sequential dissimilarity protein sequences makes the prediction of the AFPs a difficult task.  Previous sequence-based AFP predictors make use of the whole protein sequence.  In this work we propose a novel concept of the localized analysis of AFP sequences.  Extensive experiments on a number of standard datasets have been conducted.  The proposed RAFP-Pred approach has shown to perform better compared to the previous predictors such as AFP-PseAAC, AFP\_PSSM, AFP-Pred and iAFP.  The Weka model of the proposed approach have been made publicly available for benchmarking purposes (https://goo.gl/3i7gQD).  Our favorable results suggest further explorations in this direction.  For instance a more extensive segmentation could be a possible area of future research.

\section{Acknowledgement}

The authors would like to thank University of Western Australia (UWA), Pakistan Air Force - Karachi Institute of Economics and Technology (PAF-KIET), and Iqra University (IU), for providing the necessary support towards conducting this research and the anonymous reviewers for their important comments.

\bibliographystyle{IEEEtran}
\bibliography{RAFP_Pred}

\end{document}